\documentclass[sn-basic,Numbered]{sn-jnl}

\usepackage{graphicx}%
\usepackage{manyfoot}%
\usepackage{enumitem} 
\usepackage[many]{tcolorbox}

\newtcolorbox{fancybox}[1][]{
  enhanced,
  attach boxed title to top center={yshift=-3mm,yshifttext=-1mm},
  colback=blue!5!white,
  colframe=blue!25!black,
  colbacktitle=blue!25!black,
  fonttitle=\bfseries,
  title=#1,
  boxed title style={size=small, colframe=blue!45!black, colback=blue!45!white},
  drop fuzzy shadow,
  width=\textwidth,
  breakable
}

\newcommand{\sectopic}[1]{\vspace{0.2em}\par\noindent{\textit{\bfseries #1}}}
\newcommand{\iMove}{ActApp}

\begin{document}

\title{Advancing Requirements Engineering through Generative AI: Assessing the Role of LLMs}


\author*[1]{\fnm{Chetan} \sur{Arora}}\email{chetan.arora@monash.edu}

\author[1]{\fnm{John} \sur{Grundy}}\email{john.grundy@monash.edu}

\author[2]{\fnm{Mohamed} \sur{Abdelrazek}}\email{mohamed.abdelrazek@deakin.edu.au}

\affil*[1]{
\orgname{Monash University}, 
\state{Vic}, \country{Australia}}

\affil[2]{
\orgname{Deakin University}, 
\state{Vic}, \country{Australia}}

\abstract{Requirements Engineering (RE) is a critical phase in software development including the elicitation, analysis, specification, and validation of software requirements. Despite the importance of RE, it remains a challenging process due to the complexities of communication, uncertainty in the early stages and inadequate automation support. 
In recent years, large-language models (LLMs) have shown significant promise in diverse domains, including natural language processing, code generation, and program understanding. This chapter explores the potential of LLMs in driving RE processes, aiming to improve the efficiency and accuracy of requirements-related tasks. We propose key directions and SWOT analysis for research and development in using LLMs for RE, focusing on the potential for requirements elicitation, analysis, specification, and validation. We further present the results from a preliminary evaluation, in this context.
}

\maketitle

\vspace*{-1em}
\section{Introduction}~\label{sec:introduction}
\vspace*{-.5em}

Requirements Engineering (RE) is arguably the most critical task in the software development process, where the needs and constraints of a system are identified, analyzed, and documented to create a well-defined set of requirements~\cite{Pohl:10}. 
Organizations and project teams often overlook or do not understand the significance of RE and its impact on project success~\cite{vanLamsweerde:09}. 
Some underlying reasons for the lack of effort and resources spent in RE include (i)~time, budget and resource constraints, (ii)~inadequate training and skills, (iii)~uncertainty and ambiguity in early stages, which teams consider as challenging, causing them to cut corners in the RE process; (iv) inadequate tools and automation support~\cite{Arora:15a}, and (v) emphasis on an implementation-first approach instead~\cite{laplante2022requirements}. These lead to significant challenges in the later stages of development as issues related to inconsistent, incomplete and incorrect requirements become increasingly difficult to resolve, resulting in increased development costs, delays, and lower-quality software systems~\cite{Pohl:10}. 

In this chapter, we contend that the recent advances in Large-Language Models (LLMs)~\cite{Jurafsky:20} might be revolutionary in addressing many of these RE-related challenges noted above, though with some caveats. 
LLMs are advanced AI models designed to process and generate human language by learning patterns and structures from vast amounts of text data. These models have made significant strides in natural language processing (NLP) tasks and are particularly adept at handling complex language-based challenges. LLMs including OpenAI's Generative Pre-trained Transformer (GPT) series and Google's Bidirectional Encoder Representations from Transformers (BERT)~\cite{Devlin:18} and LaMDA~\cite{thoppilan2022lamda}, learn to comprehend and generate human language by predicting the most probable next word in a given sequence, capturing the probability distribution of word sequences in natural language (NL).
%
OpenAI's ChatGPT\footnote{\url{https://chat.openai.com/}} and Google's Bard\footnote{\url{https://bard.google.com/}}, built on the advancements of the LLMs, are examples of chatbot platforms designed to facilitate interactive and dynamic text-based conversations. When a user provides input to ChatGPT or Bard, the model processes the text and generates a contextually appropriate response based on the patterns learned during the training process.
%

A large majority of requirements are specified using natural language (NL). LLMs thus have the potential to be a `game-changer' in the field of RE. This could be by automating and streamlining several crucial tasks and helping to address many of the RE challenges mentioned earlier. With the focus on automated code generation using LLMs, delivering concise and consistently unambiguous specifications to these models (as prompts) becomes paramount. This underscores the ever-growing significance of RE in this new era of generative AI-driven software engineering.
This chapter explores the potential of LLMs to transform the RE processes. We present a SWOT (Strengths, Weaknesses, Opportunities and Threats) analysis for applying LLMs in all key RE stages, including requirements elicitation, analysis, and specification. 
We also discuss examples from a preliminary evaluation as motivation for using LLMs in all RE stages.

\sectopic{Preliminary Evaluation Context.} 
We performed a preliminary evaluation on a real-world app (pseudonym~\iMove), encouraging patients with type-2 diabetes (T2D) to remain active. To ensure that the app is effective, engaging, and personalized, the \iMove~team implemented a machine learning (ML) model in the background to learn from user behaviour and preferences and suggest appropriate reminders and activities. The team has a mix of experienced engineers and an ML scientist (with little understanding of RE). Our preliminary evaluation and the examples in the chapter are done using ChatGPT (GPT-3.5).


\sectopic{Structure.} Section~\ref{sec:approach} provides an overview of our vision of the role of LLMs in RE process. Sections~\ref{sec:elicitation}, \ref{sec:specification}, \ref{sec:Analysis} and \ref{sec:Validation} cover the four major RE stages, i.e., elicitation, specification, analysis and validation, respectively.
Section~\ref{sec:evaluation} presents our preliminary evaluation results. Section~\ref{sec:lessons} covers the lessons learned, and Section~\ref{sec:conclusion} concludes the chapter.

\begin{figure}[!t]
\centering
\includegraphics[width=0.8\textwidth]{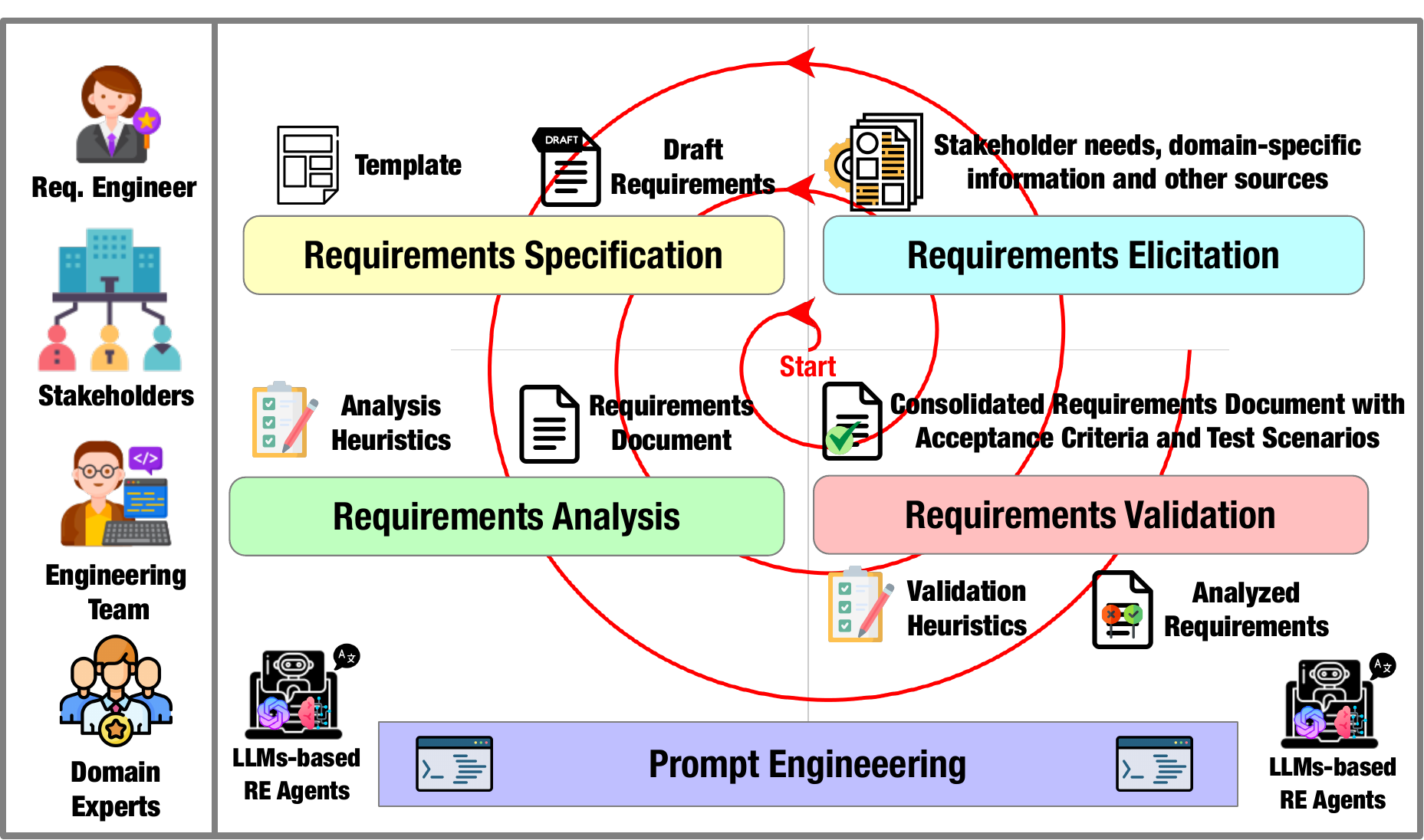}
\caption{LLMs-driven RE Process Overview.}
\label{fig:overview}
\vspace*{-1em}
\end{figure}

\section{LLMs-driven RE Process}~\label{sec:approach}
Fig.~\ref{fig:overview} provides an overview of our vision of an LLMs-driven RE process (an adaptation of RE process by Van Lamsweerde~\cite{vanLamsweerde:09}). The RE process can be broadly divided into four stages: requirements elicitation (domain understanding and elicitation), specification (specification and documentation), analysis (evaluation and negotiation), and validation (quality assurance). We note that the exact instantiation and contextualization of LLMs in RE will depend on the problem domain and the project. For instance, implementing the LARRE framework for \iMove~might be different from a safety-critical system. We, in this book chapter, provide a broad perspective on the role of LLMs in RE, which should be applicable to a wide range of projects, as the RE stages discussed are common and can be generalized across domains and systems, with finer refinements required in some cases. 

LLMs can be employed differently for automating RE tasks, e.g., as they have been successfully applied for ambiguity management~\cite{Ezzini:22}. In this chapter, we specifically focus on prompting by requirements analysts or other stakeholders directly on generative AI agents, e.g., ChatGPT or fine-tuned LLMs RE agents built on top of these agents. One would generate multiple agents based on LLMs for interaction (via prompting) with the stakeholders (e.g., domain experts, engineering teams, clients, requirements engineers and end users) and potentially with each other for eliciting, specifying, negotiating, analysing, validating requirements, and generating other artefacts for quality assurance.
Prompting is a technique to perform generative tasks using LLMs~\cite{hariri2023unlocking}. Prompts are short text inputs to the LLM that provide information about the task the LLM is being asked to perform. 
Prompt engineering is designing and testing prompts to improve the performance of LLMs and get the desired output quality. Prompt engineers use their knowledge of the language, the task at hand, and the capabilities of LLMs to create prompts that are effective at getting the LLM to generate the desired output~\cite{white2023prompt}. 
Prompt engineering involves selecting appropriate prompt patterns and prompting techniques~\cite{white2023prompt}. Prompt patterns refer to different templates targeted at specific goals, e.g., Output Customization pattern focuses on tailoring the format or the structure of the output by LLMs. Other generic templates, include formatting your prompts consistently in ``Context, Task and Expected Output'' format. For instance, one can use a \emph{persona} for output customization, wherein the agent plays a certain role when generating the output, e.g., the patient in \iMove. Prompting technique refers to a specific strategy employed to get the best output from the LLM agents. Some of the well-known prompting techniques include zero-shot prompting~\cite{pan2023preliminary}, few-shot prompting~\cite{ma2023fairness}, chain-of-thought prompting~\cite{wei2023chainofthought} and tree-of-thought prompting~\cite{yao2023tree}. In this context, prompt engineering combinations must be empirically evaluated in RE for different systems and domains. In each section, we explore the role of LLMs in each RE stage with a SWOT analysis. The insights for the SWOT analysis were systematically derived from a combination of our direct experiences with LLMs, feedback gathered from practitioner interactions, and our preliminary evaluation.

\tcbset{
    frame code={}
    center title,
    left=0pt,
    right=0pt,
    top=0pt,
    bottom=0pt,
    colback=green!10!,
    colframe=white,
    width=\textwidth,
    enlarge left by=0mm,
    boxsep=5pt,
    arc=0pt,outer arc=0pt,
    }
    
\section{Requirements Elicitation}~\label{sec:elicitation}
\vspace*{-2em}
\subsection{Elicitation Tasks} Requirements Elicitation 
encompasses pre-elicitation groundwork (as-is analysis and stakeholder analysis) and core elicitation activities with stakeholders (interviews and observations)~\cite{Pohl:10}. The main objective is to identify and document the project information, system needs, expectations, and constraints of the solution under development. The key tasks in elicitation include domain analysis, as-is analysis, stakeholders analysis, feasibility analysis, and conducting elicitation sessions with the identified stakeholders using techniques such as interviews and observations.  
While the elicitation process is methodical, it is inherently dynamic, often necessitating iterative sessions as requirements evolve and new insights emerge from stakeholders. Requirements elicitation is also intensely collaborative, demanding constant feedback and validation from diverse stakeholders to ensure clarity and alignment.
Some prevalent challenges associated with requirements elicitation involve the lack of domain understanding~\cite{sawyer1999capturing}, unknowns (i.e., known and unknown unknowns)~\cite{sutcliffe2013requirements}, communication issues due to language barriers or technical jargon~\cite{Arora:17}, and lack of a clear understanding of what needs to be built in early stages~\cite{gorschek2006requirements}. In addition, the current elicitation techniques fall short in human-centric software development, i.e., ensuring adequate representation from all potential user groups based on their human-centric factors, such as age, gender, culture, language, emotions, preferences, accessibility and capabilities~\cite{Hidellaarachchi22}. External influences, such as evolving legal stipulations and legal compliance, also play a pivotal role in shaping the elicitation process.  Furthermore, with the rapidly advancing technological landscape, the existing elicitation processes often fail to capture the system requirements precisely, e.g., in the case of AI systems, bias, ethical considerations, integration of undeterministic AI components in larger software systems~\cite{ahmad2023requirements}.

\subsection{Role of LLMs}
LLMs can address numerous key challenges in the elicitation phase, including domain analysis. 
LLMs can rapidly absorb vast amounts of domain-specific literature, providing a foundational structuring and acting as a proxy for domain knowledge source~\cite{luitel2023using}. They can assist in drawing connections, identifying gaps, offering insights based on the existing literature, and based on automated tasks such as as-is analysis, domain analysis and regulatory compliance. In addition to stakeholder communication, leveraging LLMs would require other inputs such as existing domain or project-specific documentation (e.g., fine-tuning LLMs) and regulations (e.g., GDPR). While LLMs have access to the domain knowledge, it is difficult to replace domain specialists' intuition, experience, and expertise. For example, in~\iMove~the nuanced understanding of how specific exercises influence a patient's glucose or hormonal levels rests with medical professionals such as endocrinologists, who are irreplaceable in RE.

 LLMs help identify unknowns by analyzing existing documentation and highlighting areas of ambiguity or uncertainty. LLMs can help with the completion or suggest alternative ideas that the requirements analysts might have otherwise missed, drawing on their large corpus of training data and connections. LLMs can assist with translating complex technical jargon into plain language and aiding stakeholders from different linguistic backgrounds, e.g., translating medical terminology in \iMove~for requirements analysts or translating domain information from one language to another. 
 
 LLMs play a vital role in human-centric RE. They can analyze diverse user feedback, like app reviews, ensuring all user needs are addressed. LLMs can also simulate user journeys considering human-centric factors, but this necessitates resources such as app reviews, persona-based use cases, and accessibility guidelines. For emerging technologies, LLMs need regular updates, a challenging task since automated solutions might be affected by these updates. 
 The use of LLMs in requirements elicitation also warrants ethical scrutiny. LLMs may introduce or perpetuate biases as they are trained on vast internet data. Ensuring the ethical use of LLMs means avoiding biases and guaranteeing that the stakeholders' inputs are managed according to the data privacy and security guidelines. LLMs output should be viewed as complementary to human efforts. Requirements analysts bring domain expertise, cultural awareness, nuanced understanding, and empathetic interactions to the table, ensuring that software requirements cater to the diverse and evolving needs of end-users. This synergy of humans and generative AI is crucial in human-centric software development.

\begin{tcolorbox}[colback=red!10!,
    colframe=white, breakable]
\textbf{Example Prompt for requirements generation.}
\textit{I am developing an app called \iMove. \iMove~is a real-time application for T2D patients to ensure an active lifestyle. The app gives timely reminders for working out, health \& disease management. Act and respond as an~\iMove~user with the persona provided below in JSON format. The main aim is to elicit the requirements from your perspective. The generated requirements should each be associated with a unique id, and rationale.  \\
\{``persona":\{ ``name": ``Jane Doe", 
``age": ``65",
``gender": ``Female", 
``location": ``Canada",
``occupation": ``Retired", ``medical info": $\ldots$, 
``lifestyle": $\ldots$, ``goals": $\ldots$ 
``work": ``sedentary", 
``challenges": $\ldots$ \}
\}
}
\end{tcolorbox}

\sectopic{Example.}
{For the \iMove, LLMs are used to gather information from various stakeholders, including patients and carers. The agent can conduct virtual interviews with the stakeholders (for a given persona, as exemplified below), asking targeted questions to identify their needs, preferences, and concerns. For instance, the agent may inquire users about desired features and data privacy concerns.
Additionally, LLMs can analyze and synthesize information from online forums, social media, reviews from similar apps, and research articles on disease management to extract insights into common challenges patients face and best practices for care. This information can generate preliminary requirements (e.g., R1 and R2 below), which can be refined in later stages.}

\vspace*{-0.5em}
\begin{tcolorbox}
\textbf{\iMove~Example Information and Early Requirements.}\\
\textit{Key stakeholders (identified based on initial app ideas):} Patients, carers, app developers, ML scientists, and healthcare professionals, e.g., endocrinologists.\\
\textit{R1. The patients should receive a notification to stand up and move around if they have been sitting for long.}\\
\textit{R2. The patients should not receive notifications when busy.}
\end{tcolorbox}
\vspace*{-0.5em}

\begin{fancybox}[SWOT Analysis: LLMs for Requirements Elicitation]
[{Strengths}]
\begin{itemize}[leftmargin=*]
    \item \emph{Interactive Assistance}: Can actively assist in elicitation, asking probing questions and generating diverse potential requirements based on initial inputs -- leading to uncovering unknowns.
    \item \emph{Efficient Data Processing}: Facilitate round-the-clock elicitation, rapidly processing large volumes of elicitation data in varied formats.
    \item \emph{Domain Knowledge}: Can rapidly absorb and understand domain-specific literature and automate tasks based on the absorbed literature.
    \item \emph{Assisting Multilingual and Multicultural Stakeholders}: Can accurately translate complex technical jargon into plain language and aid stakeholders' communication even with diverse backgrounds.
\end{itemize}

[{Weaknesses}]
\begin{itemize}[leftmargin=*]
    \item \emph{Lack of Empathy and Nuance}: Do not possess human empathy and might miss out on emotional cues or implicit meanings.
    \item \emph{Lack of Domain Expertise}: While LLMs understand domain knowledge, they cannot replace the intuition and experience of domain experts.
    \item \emph{Misinterpretation Risks}: The potential for misinterpreting context or over-relying on existing training data without considering unique project nuances.
\end{itemize}

[{Opportunities}]
\begin{itemize}[leftmargin=*]
    \item \emph{Real-time Documentation and Processing}: Can document requirements and analyze feedback in real time, ensuring thoroughness and accuracy.
    \item \emph{Human-centric Elicitation}: By analyzing diverse user feedback, LLMs can ensure all user needs are considered, promoting a holistic approach to elicitation.
\end{itemize}

[{Threats}]
\begin{itemize}[leftmargin=*]
    \item \emph{Over-reliance and Trust Issues}: Excessive dependence might lead to missing human-centric insights, and some stakeholders might hesitate to engage with AI.
    \item \emph{Data Security and Privacy Concerns}: Eliciting requirements via LLMs could raise data confidentiality issues, especially with sensitive information  (e.g., in public LLMs-based agents like ChatGPT and Bard).
    \item \emph{Potential Biases}:  May inadvertently introduce or perpetuate biases in the elicitation process if trained on biased data or past flawed projects.
    \item \emph{Regular Updates and Compatibility}: Given the stochastic nature of LLMs, the regular updates might lead to technical issues and inconsistency in project requirements. On the other hand, outdated LLMs are suboptimal for RE.
\end{itemize}

\end{fancybox}

\section{Requirements Specification}~\label{sec:specification}
\vspace*{-2em}
\subsection{Specification Tasks} Requirements Specification translates the raw, elicited requirements information into structured and detailed documentation, serving as the system design and implementation blueprint. LLMs can contribute to this process by helping to
generate well-structured requirements documents that adhere to established templates and guidelines, e.g., the `shall' style requirements, user story formats, EARS template~\cite{mavin2009easy}, or specific document templates, e.g., VOLERE~\cite{robertson2000volere}. Given a project's context, the informal NL requirements need to be converted into structured specifications -- both what the system should do (functional requirements) and the quality attributes or constraints the system should possess (non-functional requirements). 

Requirements analysts must maintain consistency in terminology and style throughout the document to enhance readability and clarity. In this stage, requirements can be prioritized considering stakeholder needs, project constraints, and strategic objectives. This phase is exacting, as ambiguities or errors can lead to significant project delays and escalated costs in later stages. Moreover, it is essential to balance the level of detail (too granular or too abstract) and ensure that non-functional requirements like security and usability are adequately addressed and not sidelined. Additional tasks such as generating requirements glossary, examples and rationale, and developing user personas to ensure that the human-centric aspects are duly covered are often performed during or immediately after requirements specification.

\subsection{Role of LLMs}
LLMs can streamline the specification process. The unstructured requirements from the elicitation stage can be automatically formatted into structured templates like EARS or user stories (see the example prompt below for EARS and the example for user stories). They can further assist in categorizing requirements into functional and non-functional and classifying NFRs like performance, ethical requirements, and usability. LLMs can automate other tasks during specification, e.g., generating glossary, rationale and examples, developing personas~\cite{zhang2023personagen}. Another advantage of LLMs is their ability to cross-check requirements against existing standards, regulatory guidelines, or best practices. For a health-focused app like \iMove,  LLMs can ensure alignment with health data privacy standards and medical device directives.

LLMs can also suggest requirements prioritization by analyzing technical dependencies, project goals, and historical data. However, generating requirements prioritization requires several SE roles and deep-seeded expertise. Hence, the results produced by LLMs might be inaccurate.
On similar lines, while LLMs can enhance the speed and consistency of specification, there is a risk of `over-automation', i.e., overlooking some crucial aspects or over-trusting the requirements produced by LLMs. For instance, determining the criticality of specific NFRs—like how secure a system needs to be or how scalable—often requires human expertise. LLMs can aid the process, but decisions should be validated by domain experts familiar with the project context. Similarly, for compliance issues, it is essential to have domain experts validate the results.

\begin{tcolorbox}[colback=red!10!,
    colframe=white, breakable]
\textbf{Example Prompt.} 
Using the EARS template defined by the BNF grammar below, generate the $<$requirement$>$ from the unformatted requirement - ``The patients should not receive notifications when busy.'' 

$<$requirement$>$ ::= $<$ubiquitous$>$ $|$ $<$event-driven$>$ $|$ $<$state-driven$>$ $|$ $<$optional$>$ $|$ $<$unwanted$>$

$<$ubiquitous$>$ ::= ``The system shall $<$action$>$.'' \\
$<$event-driven$>$ ::= ``When $<$event$>$, the system shall $<$action$>$.''\\
$<$state-driven$>$ ::= ``While $<$state$>$, the system shall $<$action$>$.''\\
$<$optional$>$ ::= ``The system shall $<$action$>$.''\\
\hbox{$<$unwanted$>$::= ``The system shall $<$preventive-action$>$ to $<$unwanted-outcome$>$.''}\\

$<$action$>$ ::= $<$verb-phrase$>$\\
$<$event$>$ ::= $<$noun-phrase$>$\\
$<$state$>$ ::= $<$noun-phrase$>$\\
$<$preventive-action$>$ ::= $<$verb-phrase$>$\\
$<$unwanted-outcome$>$ ::= $<$noun-phrase$>$\\
$<$verb-phrase$>$ ::= ``a verb phrase''\\
$<$noun-phrase$>$ ::= ``a noun phrase''\\
\textbf{Example Output:} ``When patient is driving, \iMove~shall not send notifications.''
\end{tcolorbox}
\vspace*{-0.5em}

\sectopic{Example.}
In \iMove, the LLMs can generate refined requirements as user stories (desired by \iMove~team members). The requirements document may include sections such as an introduction, a description of \iMove~stakeholders, a list of functional and non-functional requirements, a list of \iMove~features with priorities, and any constraints or assumptions related to the development process. For non-functional requirements, such as data privacy for patients' health information, LLMs can cross-reference with regulations, e.g., HIPAA or GDPR to ensure compliance~\cite{Abualhaija:22}.

\vspace*{-0.5em}
\begin{tcolorbox}
\textbf{\iMove~Example (as user story for functional requirements).}\\
\textit{R1.1. As a user, I want to receive a notification to move if I have been sitting for 60 minutes, so that I will be active.}\\
\textit{R1.2. As a carer, I want \iMove~to notify me if the patient continues to remain inactive after receiving a notification to move, so that I can intervene.}\\
\textit{NFR1.1: The app shall encrypt all data during transmission and storage to ensure patient privacy and comply with GDPR guidelines.}
\end{tcolorbox}
\vspace*{-0.5em}

We note that for SWOT analysis in subsequent phases, we attempt to reduce overlap (to the best extent possible). For instance, almost all threats from elicitation are also applicable for specification.

\begin{fancybox}[SWOT Analysis: LLMs for Requirements Specification]
[{Strengths}]
\begin{itemize}[leftmargin=*]
    \item \emph{Automation}: Can streamline converting raw requirements into structured formats, such as EARS or user stories. Can generate additional artefacts, e.g., glossaries and personas, from converted requirements and domain information.
    \item \emph{Compliance Check}: Can cross-reference requirements against standards and regulatory guidelines, ensuring initial compliance.
    \item \emph{Requirement Classification}: Can categorize requirements into functional and non-functional, further classifying them.
    \item \emph{Initial Prioritization}: Can suggest requirement prioritization based on dependencies, project goals, and historical data.
\end{itemize}

[{Weaknesses}]
\begin{itemize}[leftmargin=*]
    \item \emph{Depth of Domain Expertise}: While LLMs have vast knowledge, they might not fully capture the nuances of specialized domains.
    \item \emph{Over-Automation Risk}: Sole reliance on LLMs might lead to overlooking crucial requirements or business constraints.
    \item \emph{Ambiguity Handling}: May sometimes struggle with ambiguous or conflicting requirements, necessitating human intervention.
\end{itemize}

[{Opportunities}]
\begin{itemize}[leftmargin=*]
    \item \emph{Continuous Feedback}: Can aid in real-time documentation and specification updates as requirements evolve.
    \item \emph{Human-Centric Focus}: Can help maintain a human-centric outlook in the specification stage by generating alternate requirements for different user groups.
\end{itemize}

[{Threats}]
\begin{itemize}[leftmargin=*] 
    \item \emph{Ambiguities in Structured Requirements}: Can generate requirements in specific formats. However, the generated requirements can have unintentional ambiguities (if the model is not fine-tuned adequately) or other quality issues (e.g., inconsistencies due to the limited `memory' of LLMs).
    \item \emph{Over-specification}: Known to be verbose~\cite{zheng2023judging}, which can easily lead to over-defined requirements, and consequently lead to a rigid system design.
    \item \emph{Missing Non-functional Requirements}: Non-functional (unlike functional) requirements rely on a deeper understanding of the system's context, which LLMs might miss or inadequately address.
\end{itemize}
\end{fancybox}

\section{Requirements Analysis}~\label{sec:Analysis}
\vspace*{-2em}
\subsection{Analysis Tasks}
Requirements analysis focuses on understanding, evaluating, and refining the gathered requirements to ensure they are of high quality, i.e., coherent, comprehensive, and attainable, before moving to the design and implementation stages. An integral component of this phase is the automated evaluation of requirements quality. This includes addressing defects like ambiguity resolution, ensuring consistency, and guaranteeing completeness. Deficiencies in this phase can affect subsequent artefacts, leading to project delays, budget overruns, and systems misaligned with stakeholder expectations.
The main challenges of NL requirements are ambiguity, incompleteness, inconsistency and incorrectness, which lead to misinterpretations, untestable requirements, untraced requirements to their origin, no consensus among stakeholders on what needs to be built, and conflicting requirements. Constantly evolving requirements further exacerbate all these issues. At times, documented requirements or underlying assumptions might inadvertently overlook potential risks or dependencies. In such instances, it becomes crucial to identify these risks and introduce new requirements as countermeasures. Analysis of requirements, for instance, getting an agreement on conflicting requirements requires negotiation. Negotiation is the key to resolving all conflicts, and the stakeholders converge on a unified set of requirements. From a human-centric RE perspective, the analysis stage must prioritize users' emotional, cultural and accessibility needs. This entails scrutinizing user feedback for inclusivity, vetting ethics and bias concerns—especially in AI-based software systems~\cite{ahmad2023requirementsSLR}—and analyzing requirements against prevailing accessibility guidelines.

\subsection{Role of LLMs}

LLMs come into play as powerful tools to automate the quality evaluation process:

\begin{enumerate}

\item \textbf{Automated Evaluation for Quality Assurance:} LLMs can automatically assess the quality of requirements, flagging any ambiguities, vague terms, inconsistencies, or incompleteness, and highlight gaps or overlaps.

\item \textbf{Risk Identification and Countermeasure Proposal:} LLMs, when equipped with domain knowledge, can identify potential risks associated with requirements or their underlying assumptions. Drawing from historical data or known risk patterns, LLMs can suggest new requirements that act as countermeasures to mitigate those risks, ensuring system design and operation robustness.

\item \textbf{Conflict Resolution and Negotiation:} By identifying areas of contention, LLMs can facilitate the negotiation process. Multiple LLM agents can be employed to negotiate the requirements, suggest compromises, and simulate various scenarios, helping stakeholders converge on a unified set of requirements.
    
\item \textbf{Human-centric Requirements Enhancement}: LLMs can evaluate requirements to ensure they cater to diverse user needs, accessibility standards, and user experience guidelines. LLMs can also suggest requirements that enhance the software's usability or accessibility based on user personas or feedback. Moreover, they can evaluate requirements for biases or potential ethical concerns, ensuring that the software solution is inclusive and ethically sound.

\item \textbf{Change Impact Analysis:} LLMs offer real-time feedback in requirements refinement, enhancing the efficiency of the iterative analysis and maintaining stakeholder alignment. The change impact analysis process implemented as continuous feedback cycle via LLMs ensures consistency. LLMs can further proactively predict requirements changes improving the quality of requirements.
\end{enumerate}

\begin{tcolorbox}[colback=red!10!,
    colframe=white, breakable]
\textbf{Example Prompt.} 

\emph{Context:} For the \iMove~system, we need to negotiate and prioritize requirements (FR1-FR7 and NFR1-NFR5) that ensure the system caters to the patient's health needs while maintaining usability and data privacy. 

\emph{Task:} Create two agents:
Agent1 (A1) represents the primary user (a T2D patient).
Agent2 (A2) represents the system's software architect.
A1 and A2 will negotiate and discuss FR1 - FR7 to determine a priority list. During this negotiation, A1 will focus on the user experience, health benefits, and practical needs, while A2 will consider technical feasibility, integration with existing systems, and the architectural perspective. The agents can sometimes have differing opinions, leading to a more nuanced and realistic discussion. No decisions should violate NFR1 - NFR5.

\emph{Expected Output Format:} FRs in decreasing order of priority, and include the rationale for priority order based on the negotiation outcomes between A1 and A2.
\end{tcolorbox}

\sectopic{Example.} In the context of \iMove, LLMs can (i) identify and resolve ambiguities or inconsistencies in the requirements, such as conflicting preferences of patients or unclear feature descriptions; (ii) highlight any dependencies and requisites, e.g., a secure data storage system to support medical data storage; and (iii) generate missed ethical and regulatory concerns related to data storage.

\vspace*{-0.5em}
\begin{tcolorbox}
\textbf{\iMove~Analysis Examples.}\\
Identify the missing information from R1.2 and NFR1.1 in Section~\ref{sec:specification}, wherein in R1.2 - the information on how long after the initial notification the system should wait before notifying the carer is missing, and in NFR1.1, no information about data retention and deletion were specified, with regards to GDPR. 
\end{tcolorbox}
\vspace*{-0.5em}

\begin{fancybox}[SWOT Analysis: LLMs for Requirements Analysis]
[{Strengths}]
\begin{itemize}[leftmargin=*]
    \item \emph{Automation Support}: Can automatically and efficiently assess and enhance the quality of requirements, addressing ambiguities, inconsistencies, incompleteness, potential risks and countermeasures, and conflicts.
    \item \emph{Consistency}: Unlike human analysts who might have varying interpretations or might overlook certain aspects due to fatigue or bias, LLMs provide consistent analysis, ensuring uniformity in the analysis process.
    \item \emph{Historical Data Analysis}: Can draw insights from historical project data, identifying patterns, common pitfalls, or frequently occurring issues and provide proactive analysis based on past experiences.
    \item \emph{Support Evolution and Continuous Learning}: Provide real-time feedback during iterative requirements analysis, predicting possible changes and ensuring consistency. As LLMs are exposed to more data, they can continuously learn and improve, ensuring their analysis is refined.
\end{itemize}

[{Weaknesses}]
\begin{itemize}[leftmargin=*]
    \item \emph{Lack of Nuanced Domain Understanding}: Can process vast amounts of information but might miss or get confused on subtle nuances or domain context that a human analyst would catch, leading to potential oversights.
    \item \emph{Difficulty with Ambiguities}: Struggle with inherently ambiguous or conflicting requirements, potentially leading to misinterpretations for all analysis tasks.
    \item \emph{Limited Context/Memory}: Have a limited ``window'' of context they can consider at any given time. This means that when analyzing large requirements documents as a whole, they might lose context on earlier parts of the document, leading to potential inconsistencies or oversights. They don't inherently ``remember'' or ``understand'' the broader context beyond this window, which can be challenging when ensuring coherence and consistency across the document.
\end{itemize}

[{Opportunities}]
\begin{itemize}[leftmargin=*]
    \item \emph{Continuous Refinement}: As requirements evolve, LLMs can provide real-time feedback on the quality and consistency of these requirements.
    \item \emph{Integration with Development Tools}: Can be integrated with software development environments, offering real-time requirement quality checks during the software development lifecycle.
    \item \emph{Collaborative Platforms}: Can facilitate better stakeholder collaboration by providing a unified platform for requirements analysis, negotiation, and refinement.
\end{itemize}

[{Threats}]
\begin{itemize}[leftmargin=*]
    \item \emph{Over-automation}: Risk of sidelining human expertise in favor of automated checks, potentially leading to overlooked requirements defects.
    \item \emph{Regulatory Issues}: Certain industries, domains or certification bodies might have regulatory or compliance concerns related to using LLMs for critical RE tasks.
\end{itemize}
\end{fancybox}


\section{Requirements Validation}~\label{sec:Validation}
\vspace*{-2em}
\subsection{Validation Tasks}
Requirements validation ensures that the documented requirements accurately represent the stakeholders' needs and are ready for the subsequent design and implementation stages. Validating requirements often involves intricate tasks like reviewing them with stakeholders, inspecting them for defects, ensuring their traceability to their origins (or other artefacts), and defining clear acceptance criteria and test scenarios.

The primary challenge in the validation phase revolve around ensuring the requirements are devoid of gaps due to stakeholders `real' expectations and tacit assumptions. Requirements might be interpreted differently by stakeholders, leading to potential misalignments. The dynamic nature of projects means that requirements evolve, further complicating the validation process. Occasionally, requirements or their underlying assumptions might inadvertently miss certain constraints or dependencies. This leads to further issues for validation tasks. In such cases, it is imperative to identify these gaps and refine the requirements accordingly.

\subsection{Role of LLMs}

LLMs can assist in the validation phase in several nuanced ways. 
As highlighted in the Analysis phase, LLMs can aid in the manual review and inspections by flagging potential ambiguities, inconsistencies, or violations based on pre-defined validation heuristics. LLMs can be utilized to simulate stakeholder perspectives, enabling analysts to anticipate potential misinterpretations or misalignments. For instance, by analyzing historical stakeholder feedback, LLMs can predict potential areas where clarifications might be sought from the perspective of a given stakeholder. With their ability to process vast amounts of data quickly, LLMs can assist in requirements traceability to other artefacts, e.g., design documents and regulatory codes. LLMs can further assist in formulating clear and precise acceptance criteria based on the documented requirements. They can also propose test scenarios, ensuring a comprehensive validation suite.
Furthermore, LLMs can scan the requirements to identify and flag any overlooked human-centric aspects, constraints or dependencies, ensuring a more comprehensive validation. While LLMs can facilitate most validation tasks, as noted above, a major weakness of LLMs in this context is that the validation tasks often require an overall picture of the project, domain and stakeholders' viewpoints -- it is extremely difficult for LLMs to work at that level of abstraction, which typically requires manual effort from numerous stakeholders.

\begin{fancybox}[SWOT Analysis: LLMs for Requirements Validation]
[{Strengths}]
\begin{itemize}[leftmargin=*]
\item \emph{Alternate Perspectives}: Can simulate multiple stakeholder perspectives and ensure that all requirements are vetted from different viewpoints.
\item \emph{Proactive Feedback}: Can provide real-time feedback during validation sessions, enhancing stakeholder engagement.
\end{itemize}

[{Weaknesses}]
\begin{itemize}[leftmargin=*]
\item \emph{Depth of Context Understanding}: While adept at processing text, LLMs are not able to process the tacit knowledge in RE, the domain and the business context.
\end{itemize}

[{Opportunities}]
\begin{itemize}[leftmargin=*]
\item \emph{Interactive Validation Workshops}: Can be integrated into workshops to provide instant feedback, enhancing the validation process.
\item \emph{Gap Analysis Enhancements}: Can assist in refining requirements by highlighting overlooked aspects or potential improvements.
\item \emph{(Semi-)Automated Acceptance and Testing Artefacts Generation}: Can lead to a substantial effort saving in V\&V activities and concomitantly higher quality software products by generating acceptance criteria and test scenarios.
\end{itemize}

[{Threats}]
\begin{itemize}[leftmargin=*]
\item \emph{Excessive False Positives}: Likely to generate too many potential issues, leading to an unnecessary overhead of addressing false positives and slowing down the validation process -- rendering the added value for automation moot.
\item \emph{Stakeholder Misrepresentation}: Might not accurately capture the unique concerns or priorities of specific stakeholders (when simulating stakeholder perspectives), leading to a skewed validation process.
\end{itemize}
\end{fancybox}

\begin{tcolorbox}[colback=red!10!,
    colframe=white, breakable]
\textbf{Example Prompt.} 

\emph{Context:} For the \iMove~system, we need to perform the validation on all the requirements specified in the system (FR1 - FR50) and (NFR1 - NFR28). The goal is to identify the gaps in all the requirements from three different stakholders' perspectives, the software developer, the ML scientist and the product owner. 

\emph{Task:} Imagine all three stakeholders are answering this question. For each requirement, all stakeholders will write down the gaps in the requirement based on their role, and then share it with the group. Then all stakeholders will review the inputs in the group and move to the next step.
If any expert doesn't have any gap identified or a concern they can skip the discussion on that requirement. 

\emph{Expected Output Format:} For all gaps agreed upon by all stakeholders, export the issue with the requirement id.
\end{tcolorbox}


\sectopic{Example.} In \iMove, LLMs can generate acceptance criteria. Also, LLMs can uncover gaps - in our preliminary LLMs evaluation, the \iMove~team figured it needed to comply with Australia's Therapeutic Goods Act (TGA) regulation.

\vspace*{-0.5em}
\begin{tcolorbox}[
  width=\textwidth]
\textbf{Example Acceptance Criteria.}\\
\textit{R1.1-AC1 Accurately detect when the user has been sitting for 60 continuous mins.}\\ 
\textit{R1.1-AC2 Notifications can be toggled on or off by user.}\\
\textit{R2-AC1 Accurately identifies when the user is driving.}
\end{tcolorbox}
\vspace*{-0.5em}

\vspace*{-0.5em}
\section{Preliminary Evaluation}~\label{sec:evaluation}
\vspace*{-1em}

We conducted a preliminary evaluation of LLMs in RE on a real-world system (\iMove). We note that the purpose of this evaluation was not to conduct a comprehensive assessment of LLMs in RE. Instead, we focused primarily on the feasibility of integrating LLMs into requirements elicitation. The rationale is that the applicability of LLMs to the remaining RE stages is relatively intuitive, thanks in part to the extensive history and well-established methodologies of applying NLP techniques to these stages~\cite{zhao2021natural,zamani2021machine}. Thus we deemed exploring LLMs in requirements elicitation to be the essential first step.

\sectopic{Data Collection Procedure.}
The main goal of our data collection procedure was to establish the user requirements in \iMove~and analyze the performance of ChatGPT for requirements elicitation. 
Our team had access to three \iMove~experts - the project manager, an ML scientist, and a software engineer. These experts met with a researcher, Katelyn (pseudonym), to articulate the project's focus. The meetings were part of a broader context of understanding the RE processes. ChatGPT was not mentioned to experts to avoid bias. Katelyn engaged in four two-hour meetings with the experts, where they presented an overview of the project, system users, user requirements, and software features.

We used ChatGPT to simulate the initial stages of requirements elicitation, wherein requirements engineers acquire project knowledge from stakeholders, review existing documentation, and formulate user requirements and core functionalities. The process involved four participants: Jorah and Jon, both seasoned software/requirements engineers, and Arya and Aegon, both early-stage RE and NLP research students. They were given a project overview from Katelyn and asked to start a ChatGPT session, introducing themselves as developers of the \iMove~project. Guided by the project brief, they interacted with ChatGPT to elicit user-story-style requirements over a 45-minute session. Subsequently, Katelyn examined the requirements generated by the participants using ChatGPT against the actual project requirements.



\begin{table}[h]
    \centering
    \vspace*{-1em}
\caption{Evaluation Results}
\label{tab:results}
\vspace*{-1.8em}
    \begin{tabular}{|p{0.115\linewidth}|p{0.075\linewidth}|p{0.075\linewidth}|p{0.075\linewidth}|p{0.12\linewidth}|p{0.135\linewidth}|p{0.1\linewidth}|p{0.07\linewidth}|} \hline 
        \textbf{Participant} & \textbf{Elicited} & \textbf{Full Match} & \textbf{Partial Match} & \textbf{Potentially Relevant} & \textbf{Superfluous/ Redundant} & \textbf{Precision} & \textbf{Recall} \\ \hline 
         Jorah & 14 & 11 & 1 & 0 & 2 & 82\% & 58\% \\ \hline 
         Jon & 17 & 7 & 4 & 2 & 4 & 53\% & 45\% \\ \hline 
         Arya & 14 & 3 & 2 & 1 & 8 & 29\% & 20\% \\ \hline 
         Aegon & 27 & 2 & 4 & 1 & 20 & 15\% & 20\% \\ \hline
    \end{tabular}
    
\vspace*{-1.5em}    
\end{table}

\sectopic{Results.} Overall, 20 key user requirements were identified in \iMove~by Katelyn with the experts. Katelyn mapped the requirements Jorah, Jon, Arya and Aegon elicited against these 20 requirements. Each requirement in the elicited set was categorized as a full match, partial match, or no match. It should be noted that a `full match' did not imply an exact syntactic duplication of the original requirement but rather captured its essence effectively. Likewise, a `partial match' indicated that only a part of the original requirement's essence was captured. We note that in our calculation of precision and recall, each full match is weighted as 1 true positive (TP) and each partial match is weighted as 0.5 TP.
Katelyn further classified all `no match' requirements as superfluous or potentially relevant (for further expert vetting if required). 
Table~\ref{tab:results} shows the overall results from the four participants. The results clearly show the significance of experience while using ChatGPT in this preliminary evaluation. While none of the participants could elicit most requirements, it is important to note that with a project brief and one interaction session, the experienced participants could get almost half the relevant requirements, emphasising the feasibility of LLMs for RE. 

 \section{Lessons Learned}~\label{sec:lessons}
 Our preliminary evaluation provided insights and highlighted challenges noted below.

    \sectopic{Role of Prompts and Contextual Information.} LLMs depend heavily on comprehensive prompts and the availability of contextual information to generate meaningful output. Slightly different prompts can produce very different outputs. A thorough empirical evaluation of prompt engineering is necessary for employing LLM agents.

\sectopic{Experience Matters.} Experienced requirements engineers were more successful in formulating prompts, interpreting responses, and getting quality output, despite the project background being uniform across participants. This highlights the importance of experience and training in RE teams. 

\sectopic{LLMs Capabilities.} Our preliminary evaluation underlined the capability of LLMs to discover `unknown' requirements, addressing a significant challenge in RE. We found four `potentially relevant' requirements for future stages in \iMove, which were not part of the original set, from three participants. We surmise that LLMs may assist interpreting and generating text for varied stakeholders, which can be key in reducing communication barriers inherent in diverse project teams. However, managing many `false positive' candidate requirements will require care to ensure engineers are not overloaded with many irrelevant or semi-inaccurate requirements.

\sectopic{LLM Problems.} LLMs have some inherent issues, such as systematic inaccuracies or stereotypes
in the output (influenced by the training data~\cite{borji2023categorical}), and the limited context length, e.g., ChatGPT has a limit of 32K tokens, which although enough but still might make it difficult to process large documents or maintain task context in a session. All participants reported issues with maintaining the context of \iMove~system in the evaluation session and noticed inaccuracies. 

\sectopic{Domain Understanding.} RE requires an excellent understanding of the underlying domain for eliciting and specifying correct and complete requirements. An LLM's training on specific domain knowledge may be limited and requires addressing to incorporate domain knowledge via experts, other sources, or fine-tuned LLMs. Access to large amounts of training data to fine-tune a custom LLM may be a challenge.

\sectopic{Automation Bias.} Humans often display unfounded trust in AI~\cite{akter2021algorithmic}, e.g., the LLMs-generated requirements in our case. For example, upon completing the session, Arya and Aegon displayed a remarkable degree of confidence in their elicited requirements. 

\sectopic{Security, Privacy and Ethical Issues.} Requirements are by their very nature mission-critical for software engineering and incorporate much sensitive information. Disclosure via public LLMs may result in IP loss, security breaches in deployed systems, organisational and personal privacy loss, and other concerns. 
Who `owns' requirements generated by LLMs from training data from unknown sources? 


\section{Conclusion}~\label{sec:conclusion}
In this chapter, we explored the transformative potential of LLMs at various stages of RE. Our exploration positioned that LLMs have the potential to enhance several RE tasks by automating, streamlining, and augmenting human capabilities. Their capability to simulate stakeholder perspectives, generate alternative requirements, address requirements quality, cross-reference with standards, and generate structured documentation is revolutionary. However, we also cautioned against unchecked optimism (using detailed SWOT analysis) that LLMs are not the `silver bullet' for solving all RE problems and in fact have threats of their own. Specific challenges and threats associated with their application in RE include understanding deep-seeded domain nuances, understanding the overall context, over-automation, over-specification and losing the human-centric view of requirements. The chapter further outlines lessons learned from applying LLMs in a real-world project on RE for \iMove~app for T2D patients.










\bibliography{paper}

\end{document}